\documentclass[twocolumn]{aastex63}

\RequirePackage{color}

\shortauthors{Miyoshi \& Chiba}
\shorttitle{Orbits of the dSphs in the Milky Way}

\begin{document}

\title{Long-term orbital evolution of Galactic satellites and
the effects on their star formation histories}

\correspondingauthor{Masashi Chiba}
\email{chiba@astr.tohoku.ac.jp}

\author{Takahiro Miyoshi}
\affiliation{Astronomical Institute, Tohoku University, Aoba-ku, Sendai 980-8578, Japan}

\author[0000-0002-9053-860X]{Masashi~Chiba}
\affiliation{Astronomical Institute, Tohoku University, Aoba-ku, Sendai 980-8578, Japan}

\begin{abstract}
We investigate the orbital motions of dwarf spheroidal galaxies (dSphs)
in the halo of the Milky Way (MW) to understand their possible effects on the diversity
of the star formation histories seen in these MW satellites. In this work,
we explicitly consider a time-varying gravitational potential due to
the growth of the MW's dark halo mass to calculate the long-term orbital evolutions
of the dSphs, guided with {\it Gaia} DR2 proper motions, over the past 13.5 billion years.
We find that the infall time of a satellite, defined at which the galaxy
first crosses within the growing virial radius of the MW's halo, coincides well
with the time when the star formation rate (SFR) is peaked for the sample of
classical dSphs. On the other hand, ultra-faint dSphs already finished their SF activity
prior to their infall times as already suggested in previous works, but there is a signature
that their earlier SF histories are affected by interaction with the growing
MW's halo to some extent. We also find, for classical dSphs, that the relative fraction
of stars formed after the peak of the SFR to the current stellar mass is smaller for
the smaller pericentric radius of the galaxy at its first infall.
These results suggest that the infalling properties of the dSphs into the MW
and the resultant environmental effects such as ram-pressure stripping and/or
tidal disturbance in the MW's dark halo containing hot gas play important roles
in their star formation histories.
\end{abstract}


\section{Introduction} \label{sec:intro}

Our understanding of how dwarf spheroidal galaxies (dSphs) in the Milky Way (MW) have 
formed and evolved to what we see today remains still far from being complete.
The distribution of the member stars in a color-magnitude diagram indicates that each of the MW's dSphs
shows a variety of time evolution of star formation rate (SFR) \citep[e.g.,][]{Mateo1998,Grebel1999}.
The diversity of these SF histories may be summarized into the following different groups:
(1) SF occurred only in the early phase of the dSph, leading to the dominance of
old stellar populations as old as 12 Gyrs,
(2) SF occurred mostly in the near past, as inferred from the dominance of relatively
young stellar populations as young as $\sim 5$ Gyrs,
(3) SF has been occurred over several Gyrs in the middle of the galaxy history,
and (4) the galaxy had experienced episodic SF events
\citep[e.g.,][]{Buonanno1999,Grebel2004,Tolstoy2009,deBoer2012,McConnachie2012,deBoer2014}. 

In particular, deep photometric observations of Galactic satellites with
{\it Hubble Space Telescope} ({\it HST}) by \citet{Weisz2014} clearly indicate that
the so-called classical dSphs in the MW, having a $V$-band absolute magnitude, $M_{\rm V}$,
brighter than $-8$~mag, show diverse SF histories at given $M_{\rm V}$. On the other hand,
ultra-faint dwarf galaxies (UFDs) with $M_{\rm V} > -8$~mag contain only old member stars,
whereby the SFR was peaked and quenched at early epochs
\citep{Okamoto2008,Brown2012,Brown2014,Weisz2014,Simon2019}. Stars in dSphs also show
their characteristic iron abundance as well as abundance-ratio distributions \citep[e.g.,][]{Tolstoy2004,Koch2006,Battaglia2006,Tolstoy2009,Kirby2011,Kirby2013,Ishigaki2014,Tsujimoto2015,Simon2019}.
These chemical properties of dSphs must be intimately related to the diversity of their SF histories,
but what causes this diversity is unsolved yet, including their intrinsic properties
or external effects
\citep[e.g.,][]{Mayer2006,Revaz2009,McConnachie2012,Okayasu2016,Bermejo2018,Revaz2018,Escala2018}.

One of the key ingredients that controls the SF histories in Galactic dSphs is their 
past and current environment within the halo of the MW. This includes the tidal 
effects from the gravitational field of the MW, the ram-pressure stripping of 
cold gas in their progenitor galaxies in the presence of the MW's hot halo gas, the 
photo-ionizing effect of UV light from the MW as well as from the Universe, 
and so on. Indeed, recent hydrodynamical simulations of galaxy formation in the 
framework of $\Lambda$-dominated cold dark matter theory suggest that all of these 
physical processes are actually at work on each of dark-matter subhalos falling into 
a host, MW-sized halo, which may eventually become currently observed luminous satellites \citep[e.g.,][]{Wetzel2015,Maccio2017,Frings2017,Simpson2018,Buck2019,
Genina2019,Garrison-Kimmel2019,Rodriguez2019,Fillingham2019}.

To assess these environmental effects on the SF histories of Galactic dSphs, it is 
important to investigate when and how these dSphs are falling into and eventually 
orbiting in the gravitational field of the MW's dark halo and how these orbital 
evolutions are associated with the past SF events in each of the dSphs. This approach 
is possible only when the reliable kinematical information are available for many 
of the different dSphs showing different stellar populations. Several previous works
have already suggested, based on the calculations of the orbits of Galactic dSphs, that their
close passages to the MW can be linked with the SF histories, including the peak
and the subsequent time evolution of the SFR
\citep[e.g.,][]{Sohn2007,Pasetto2011,Rocha2012,Fillingham2019,Rusakov2020}.

In this respect, {\it Gaia} DR2 has revolutionized both precisions and amounts in the astrometric data
of Galactic stars, so the estimation of the precise spatial motions of many different dSphs, 
including both classical dSphs and UFDs, is now possible. Based on the calibration of the 
proper motions from {\it Gaia} DR2, \citet{Helmi2018} calculated the orbits of the 
nine classical dSphs and one UFD as well as 75 globular clusters for three different models
of static Galactic gravitational potentials and showed the distribution of their orbital properties.
\citet{Fritz2018} further derived the {\it Gaia} DR2 proper motions of more than 39 dwarf
galaxies located out to 420~kpc from the Galactic Center and integrated their orbits
in static canonical MW potentials. Similar studies for the orbits of 17 UFDs in a static
Galactic potential are made by \citet{Simon2018} based on the calibration of
the {\it Gaia} DR2 proper motions for these satellites.
\citet{Kallivayalil2018} showed that some of these satellites are associated with
the Large Magellanic Cloud (LMC) \citep[see also][]{Yozin2015,Pardy2019,Erkal2019,Patel2020}. 

However, we note that these orbital calculations of Galactic dSphs are made
under the assumption that the Galactic gravitational field is fixed with the currently observed form.
Although this assumption is appropriate for the relatively short period of a few Gyrs,
it cannot be applied to the whole history of the MW, especially at the early stage of
the first passage of satellites to the host, MW's dark halo,
whose mass is growing through hierarchical accretion processes.
Thus, it is not fully understood yet whether the orbital motions of dSphs play a major role
in their SF histories through the relevant environmental effects. 

This work intends to relax this assumption of a static Galactic potential and explores
the long-term orbital evolutions of Galactic dSphs under the situation of a growing mass of
the MW's dark halo. Similar calculations for the past orbits of the LMC or star clusters
in a time-varying Galactic potential were made by \citet{Zhang2012} and \citet{Haghi2015}
and we adopt here the method in these previous works.

Recently, \citet{Kelley2019} performed the Phat ELVIS suite of cosmological zoom-in simulations
of MW-sized galaxies, and \citet{Fillingham2019} used these simulations
to follow the orbital evolutions of subhalos relative to their host halo and derived the infall time
at which a subhalo {\it first} crosses within the virial radius of a growing MW-sized host halo.
These subhalos were then matched with 37 Galactic satellites by comparing both in the diagram of
binding energy vs. distance from host \citep{Rocha2012}, whereby the infall time inferred
for each satellite was compared with its SF history.

Here, instead of using such extensive cosmological simulations, we adopt an analytically tractable,
time-varying mass of a MW-sized host halo as in \citet{Zhang2012} and \citet{Haghi2015} to directly integrate
the long-term orbital evolution of each of Galactic dSphs and investigate their first infall
and subsequent orbital motions in comparison with a growing virial radius of the host halo.
We then infer the possible relation of these orbital motions with their SF histories. 

The paper is organized as follows. Section 2 shows the method for the calculations of the
orbital motions of the sample dSphs over many dynamical times of the Galaxy. Section 3 is
devoted to the results of these calculations. In Section 4, we discuss the physical mechanism
behind our calculated results by comparing with recent high-resolution simulations of
galaxy formation and the conclusions are made.

For all the relevant calculations in what follows, we adopt the set of the cosmological 
parameters based on WMAP7 \citep{Larson2011}: $\Omega_m = 0.266$, $\Omega_\Lambda=0.734$
and $H_0 = 71$~km~s$^{-1}$~Mpc$^{-1}$.

\section{Method} \label{sec:method}

We simply assume that the form of the Galactic potential at each epoch is
spherically symmetric and is given by the so-called NFW profile \citep{NFW1996}
\begin{equation}
\Phi(r) = - \frac{G M_{\rm vir}}{r [ \log(1+c) - c/(1+c) ]}
           \log \left( 1 + \frac{cr}{r_{\rm vir}} \right)
\label{eq: NFW}
\end{equation}
where $M_{\rm vir}$, $r_{\rm vir}$ and $c$ denote the virial mass, virial radius and 
concentration parameter of the MW's dark halo, respectively. 

In this work, we explicitly consider the time evolution of these quantities over the past 13.5
Gyrs to follow the corresponding long-term orbital evolution of Galactic satellites. Regarding the 
growth of the MW's virial mass, $M_{\rm vir}(t)$, \citet{Wechsler2002} obtained the 
approximate formula as a function of redshift, $z$, based on their cosmological N-body 
simulations,
\begin{equation}
    M_{\rm vir} (z) = M_{\rm vir} (z=0) \exp( -2 a_c z )
\label{eq: Mvir}
\end{equation}
where $a_c$ controls the growth time scale of the MW's mass, which is given as $a_c = 0.34$. In this 
case, the growth is slow, where the epoch when the MW's mass was half of the current value 
was about 8 Gyrs ago. For the current virial mass of the MW, we set
$M_{\rm vir} (z=0) = 1.54^{+0.75}_{-0.44} \times 10^{12} M_{\odot}$ taken
from the recent measurement of the spatial motions of globular clusters based on {\it Gaia} DR2's
proper motions \citep{Watkins2019} and also consider the effect of these 1~$\sigma$ uncertainties
in the $M_{\rm vir} (z=0)$ value on the calculation of the satellites' orbits.
We note that \citet{Krumholz2012} also showed the mass accretion history of the MW's halo
and the corresponding time evolution for the MW's dark halo mass is found to be nearly the same as
that Equation~(\ref{eq: Mvir}) provides.

Given $M_{\rm vir}$ at each epoch, the virial radius is estimated as the radius within which 
the mean density of the corresponding halo is 200 times as large as the critical density of the 
Universe, $\rho_c (z)$, namely, $M_{\rm vir} = (4\pi/3) r_{\rm vir}^3 200 \rho_c$. This yields 
$r_{\rm vir} (z)$, For the time evolution of $c$, we adopt the relation, $c( M_{\rm vir}, z )$, 
given by \citet{Prada2012}.

\begin{figure*}[t!]
\centering
\includegraphics[width=120mm]{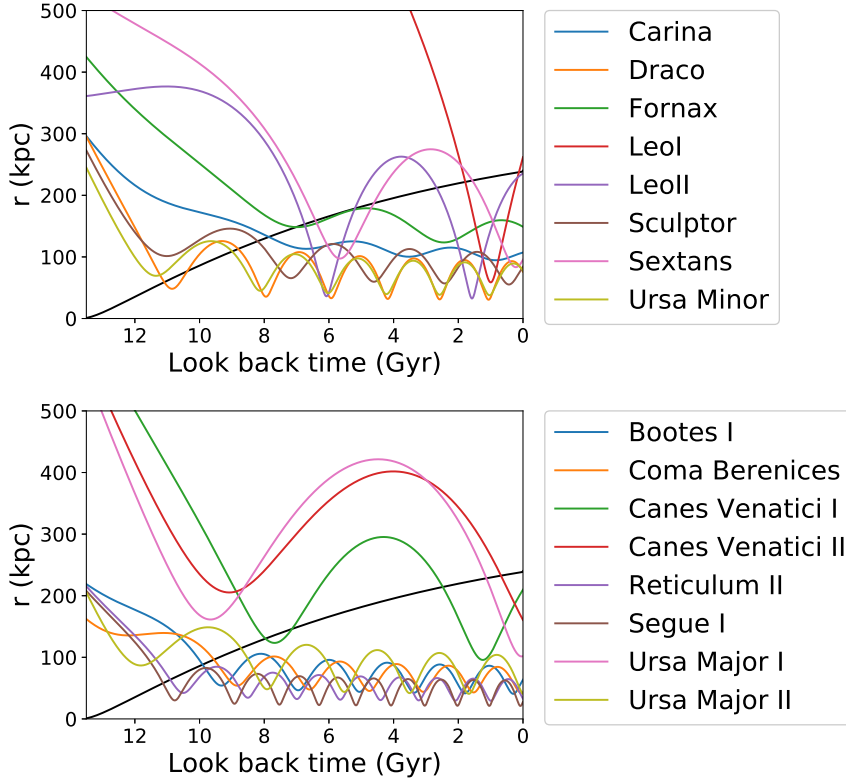}
\caption{
The time evolution of the Galactocentric distance, $r$, of the classical dSphs 
(upper panel) and UFDs (lower panel) over the past 13.5 Gyrs.
The black line in both panels shows the time 
evolution of the virial radius of the MW's dark halo, $r_{\rm vir}$, which is increasing with time
under the growing mass of the dark halo. 
}
\label{fig: orbits}
\end{figure*}

The proper motions of the sample of the dSphs that we use here are based on {\it Gaia} DR2 catalog.
Here, we adopt the list of 6D data in the work by \citet{Riley2019}, who assembled the distances,
line-of-sight velocities and proper motions for 38 Galactic satellites. Among these, we select 8 classical dSphs
(Carina, Draco, Fornax, Leo~I, Leo~II, Sculptor, Sextans and Ursa Minor)
and 8 UFDs (Bootes I, Coma Berenices, Canes Venatici I (CVn~I), Canes Venatici II (CVn~II),
Reticulum~II, Segue~I, Ursa Major I and Ursa Major II), which covers a wide range of orbital properties
as shown below and for most of which SF histories are available from the {\it HST} observations
by \citet{Weisz2014} and \citet{Brown2014}. In this sample selection for the orbit calculation,
we avoid the satellites having large uncertainties in the measured proper motions and
thus 3D velocities, such as Leo~IV and Hercules, and those being thought as LMC satellites
in recent studies \citep{Kallivayalil2018,Patel2020}.

With the above set-up, we calculate the past orbit of each satellite using {\it galpy} \citep{Bovy2015}
with a time step of 10~Myr. The Galactocentric distance of the Sun, its circular velocity and
the local solar motion for this calculation are adopted as 8~kpc, 220~km~s$^{-1}$ and (11.1, 12.24,
7.25) km~s$^{-1}$ \citep{Schoenrich2010}. We note that the virial radius is increasing with time as the 
virial mass of the MW grows. Thus, at the specific epoch, an infalling satellite first crosses within
this increasing virial radius of the MW with time, and we define this time when it occurs as ``the infall time''
hereafter denoted as $t_{\rm first~infall}$. 

We note that in these orbit calculations of Galactic satellites, the choice for the specific form of the
Galactic potential, including its spatial and time dependence, affects the resulting time-evolution of the orbits.
In particular, we assume here a spherical symmetry, $\Phi(r)$, over all the times, whereas actual accretion history
of the MW is, of course, not spherically symmetric and the growth of the MW mass is not continuous, so that
both of these affect the orbits of Galactic satellites significantly. While the consideration of all the cases
for the Galactic potential is beyond the scope of the current work, we here choose and test a different model
for $M_{\rm vir}(z)$ derived by \citet{Krumholz2012} for comparison, which is derived from
\begin{eqnarray}
\dot{M}_{\rm h,12} &=& - \alpha M_{\rm h,12}^{1+\beta} \dot{\omega} \\
\dot{\omega} &=& -0.0476 [1+z+0.093(1+z)^{-1.22}]^{2.5} \ {\rm Gyr}^{-1} ,
\label{eq: Mvir2}
\end{eqnarray}
where $\dot{}$ denotes the time derivative, $M_{\rm h,12} = M_{\rm h}/10^{12}$~$M_{\odot}$, and $M_{\rm h}$
is a halo mass, being set to $M_{\rm vir}$ in this work. It is interesting to note that the deviation of
$t_{\rm first~infall}$ from the case of adopting Equation (\ref{eq: Mvir}) is found to be generally much smaller than
1~Gyr (except Draco with the deviation of $\sim 1.5$~Gyr and Ursa Minor with $\sim 1.0$~Gyr),
whereas the uncertainties stemmed from the range of values of $M_{\rm vir} (z=0)$ are more significant.
The principal importance of the host halo mass in the orbits of the associated
subhalos is also suggested from the analysis of their orbital properties from cosmological N-body simulation
\citep{Wetzel2011}. Taking this regard into account, here we mainly consider the range of $t_{\rm first~infall}$
resulting from the choice of $M_{\rm vir} (z=0)$ within its 1~$\sigma$ uncertainties, namely,
$1.10 \times 10^{12} M_{\odot}$ to $2.29 \times 10^{12} M_{\odot}$.

\section{Results} \label{sec:results}
\subsection{Orbital properties}

Figure~\ref{fig: orbits} shows the time evolution of the Galactocentric distance, $r$,
of the 8 classical dSphs (upper panel) and 8 UFDs (lower panel).
The black lines in both panels denote the time evolution of the virial radius of
the MW's dark halo, $r_{\rm vir}(t)$.
It follows that these orbital evolutions of Galactic dSphs can be roughly divided into
three different cases: (1) the dSphs like Fornax, Leo~II, Sextans and CVn~I 
have travelled around the MW only two to three orbits, (2) the dSphs like Draco, 
Sculptor and Ursa Minor as well as the most of UFDs have fallen at early epochs and thus
have executed many orbital oscillations around the MW, (3) Leo~I is now located
beyond $r_{\rm vir}$ after having crossed within it about 2 Gyr ago and arrived at
the pericentric radius about 1 Gyr ago \citep{Sohn2007}.

\begin{figure}[t!]
\centering
\includegraphics[width=80mm]{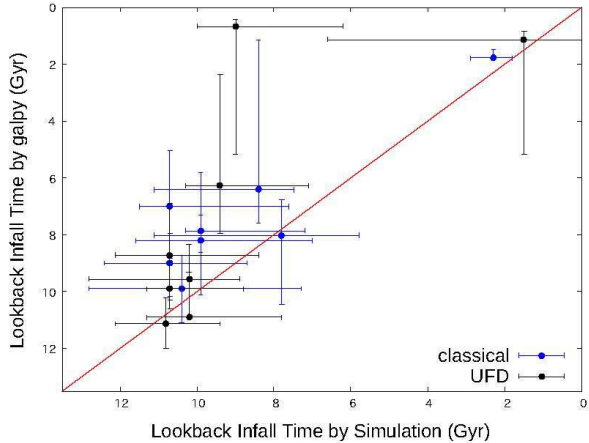}
\caption{
The comparison between the infall time in the current work (vertical axis) and that
in \citet{Fillingham2019} (horizontal axis) for classical dSphs (blue circles) and
UFDs (black circles). These infall times based on different methods and Galactic potentials
are roughly in agreement with each other, except for CVn~II having the shortest infall time
in our work. For reference, those with the 2nd and 3rd shortest infall times in our work
(located at the upper-right corner in the diagram) are Ursa Major I and Leo~I, respectively.
}
\label{fig: infall_time}
\end{figure}
For these sample satellites, we derive the infall time, $t_{\rm first~infall}$, in the
currently adopted, time-varying Galactic potential. The comparison is then made with the
infall time obtained in the simulation work of \citet{Fillingham2019}, who analyze
the orbits of subhalos in 12 MW-sized halos from the Phat ELVIS simulation \citep{Kelley2019}
and matching with the observed dSphs is made. As shown in Figure~\ref{fig: infall_time},
the currently derived infall times are roughly in agreement with those
in the work of \citet{Fillingham2019} within the uncertainties, except the case of CVn~II
perhaps due to the difference in the adopted mass models.
This suggests that the current simplified treatment of the satellites' obits given
in the Galactic potential of Equation~(\ref{eq: Mvir}) provides generally consistent results
with those based on the high-resolution simulations of evolving MW-sized dark halos.

\begin{figure}[t!]
\centering
\includegraphics[angle=-90, width=80mm]{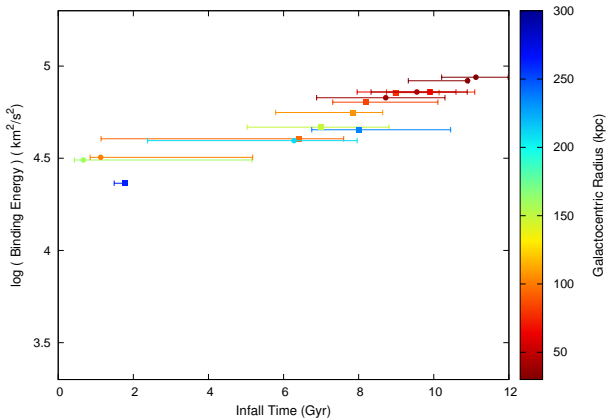}
\caption{
The comparison between the infall time and binding energy of each satellite,
where color codes are based on the Galactocentric radius.
The basic properties of increasing binding energy with increasing infall time
are consistent with those presented in \citet{Rocha2012} and \citet{Fillingham2019}.
For reference, the satellites with the 1st, 2nd and 3rd shortest infall times
are CVn~II, Ursa Major I and Leo~I, respectively.
}
\label{fig: binding}
\end{figure}

The relation between the infall time and the binding energy of each satellite is shown
in Figure~\ref{fig: binding}. The general trend of this relation is again consistent with
that obtained from the cosmological simulations of MW-sized halos
\citep{Rocha2012,Fillingham2019}, namely a satellite that has fallen earlier is
more strongly bound to the MW's halo located at the smaller Galactocentric radius.

To assess the effect of considering the growing mass of the MW on these orbits,
we also calculate the case when the Galactic potential is fixed in the form of Equation~(\ref{eq: NFW})
for all of our sample dSphs. The plots for all of these orbits are presented in Appendix. In short,
we find that while these orbits show only a simple oscillation in $r$ for a static Galactic potential,
those in the currently time-varying Galactic potential start to notably deviate from these simple oscillations
at the epochs about 4 Gyrs ago. Thus, to derive the past orbits of the dSphs
more than 4 Gyrs ago, we need to explicitly take into account the time variation of the Galactic
potential as adopted here.

\begin{figure*}[t!]
\centering
\includegraphics[width=100mm]{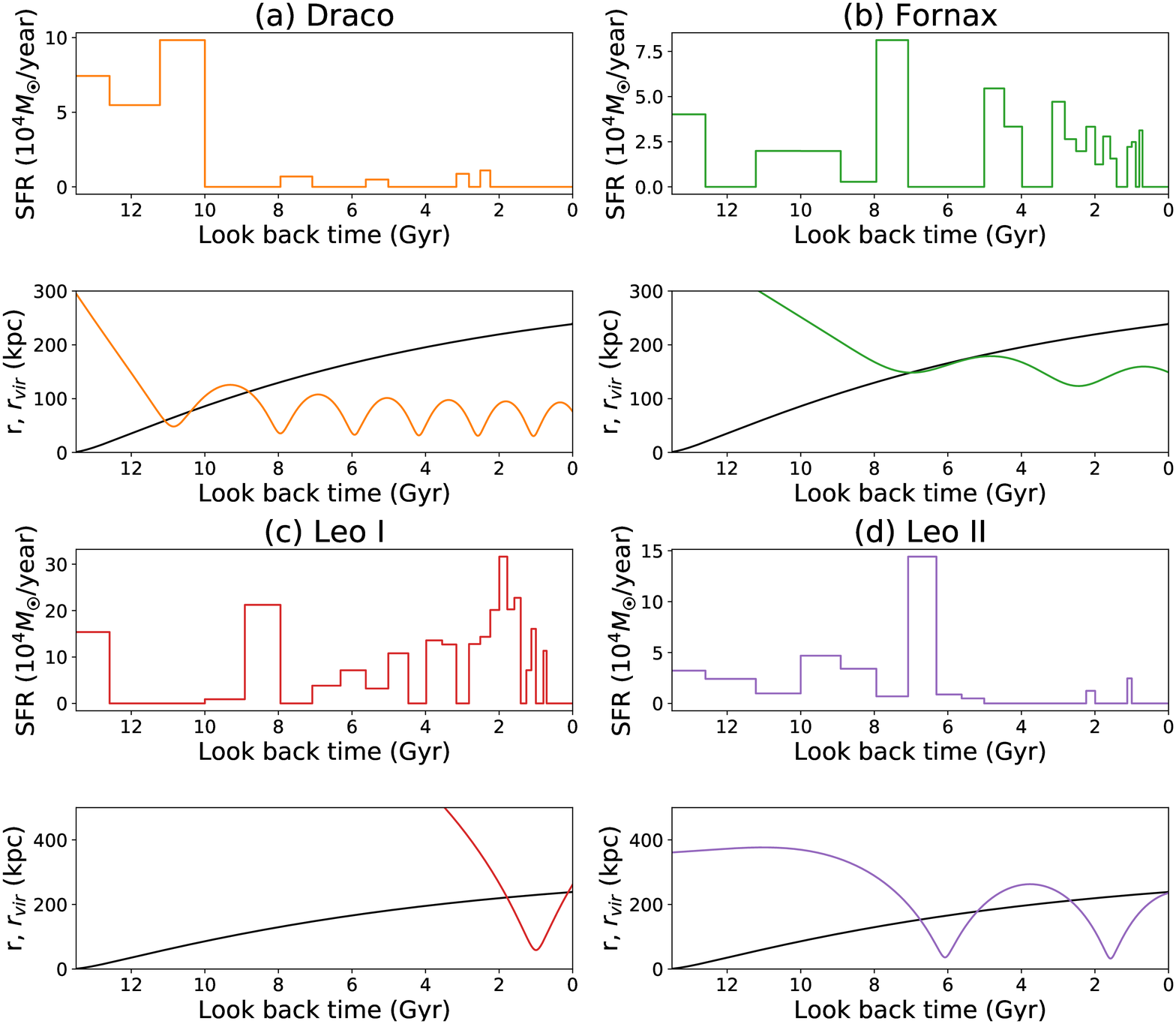}
\caption{
The comparison between the long-term orbital motions of the classical dSphs (lower panel)
and their star formation histories (upper panel) available from \citet{Weisz2014},
for (a) Draco, (b) Fornax, (c) Leo~I, and (d) Leo~II. The black solid line in each lower panel
shows the time evolution of the virial radius of the MW's halo, $r_{\rm vir}(t)$.
As is evident, the SFR is peaked at around the infall time when the satellite first crosses
within the virial radius of the MW's halo.
}
\label{fig: SFR1}
\end{figure*}
\begin{figure*}[t!]
\centering
\includegraphics[width=100mm]{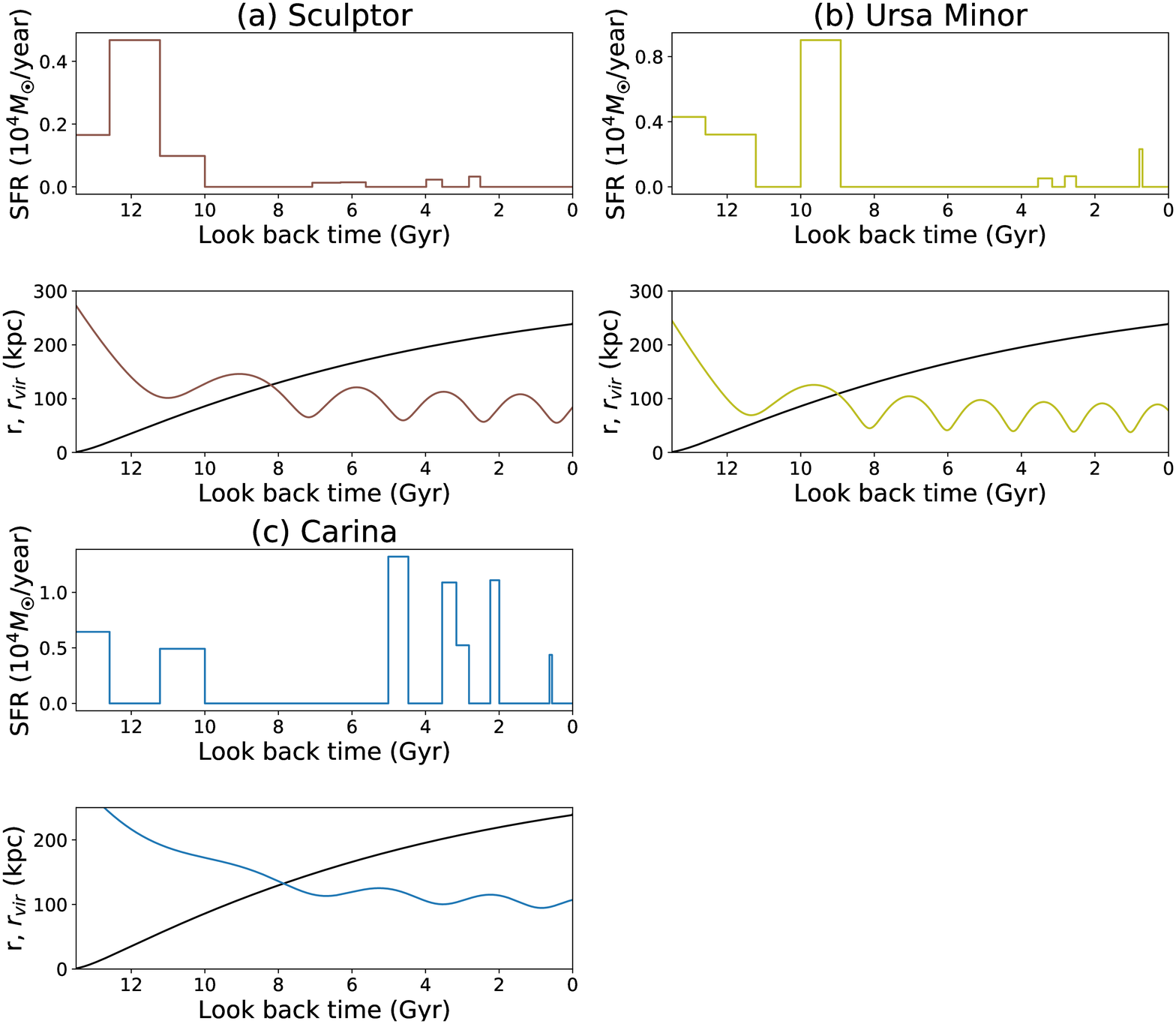}
\caption{
The same as Figure~\ref{fig: SFR1} but for the other three classical dSphs,
(a) Sculptor, (b) Ursa Minor, and (c) Carina.
}
\label{fig: SFR2}
\end{figure*}
\subsection{Comparison with star formation histories}

\begin{figure*}[t!]
\centering
\includegraphics[width=100mm]{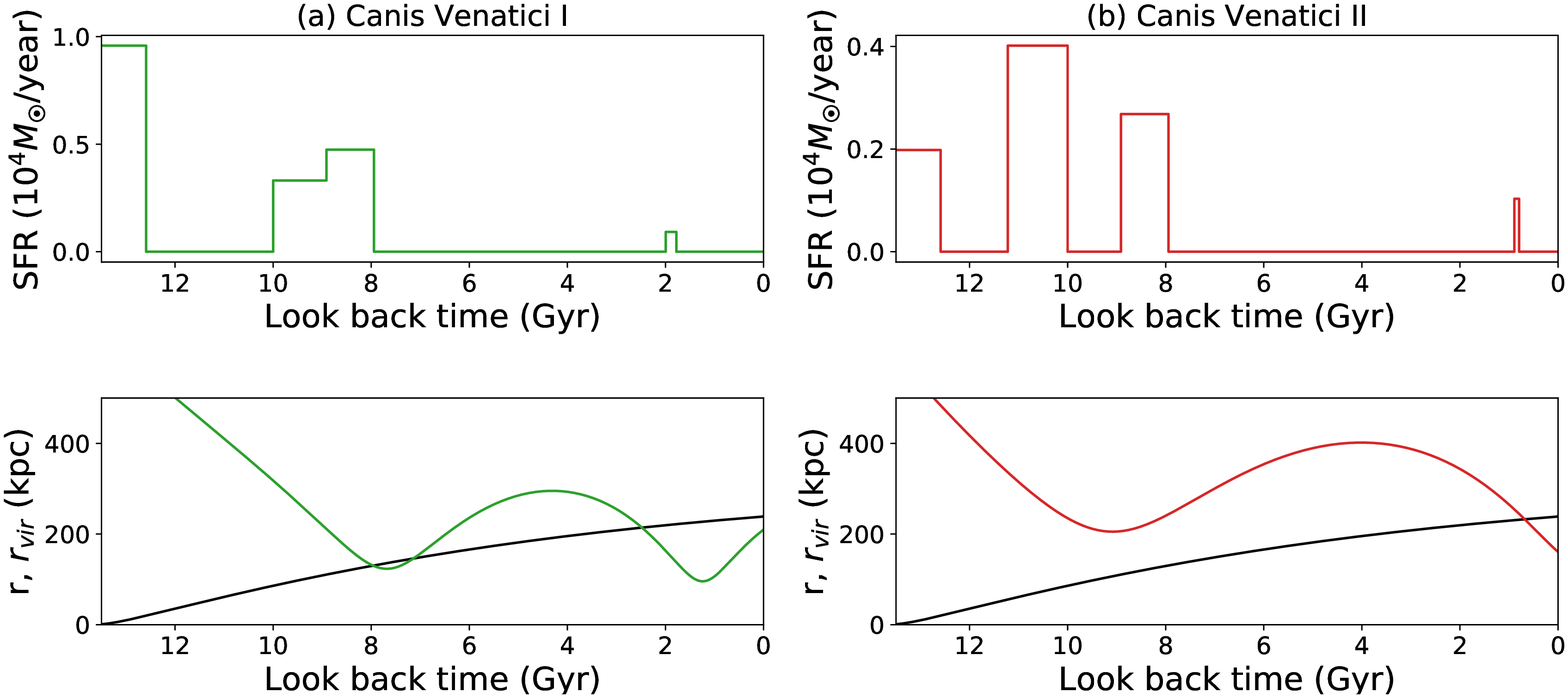}
\caption{
The same as Figure~\ref{fig: SFR1} and \ref{fig: SFR2} but for the two UFDs,
(a) CVn~I and (b) CVn~II for which star formation histories are available from \citet{Weisz2014}.
In contrast to classical dSphs, these UFDs finished star formation before the infall time.
We note that in CVn~I, the second peak of the SFR occurs nearly at the infall time,
whereas in CVn~II, the peak of the SFR occurred at around its first passage to the MW.
}
\label{fig: SFR3}
\end{figure*}

Figure~\ref{fig: SFR1} and \ref{fig: SFR2} show the comparison between these long-term orbital
motions of the seven classical dSphs (lower panel) and their SF histories (upper panel), where
the latter is expressed in terms of the differential SFR as a function of look-back time
using the cumulative SF history and total stellar mass available from the \citet{Weisz2014} work.
Figure~\ref{fig: SFR3} is the same as these figures but for two UFDs, CVn~I and CVn~II, for which
SF histories are again available from the same reference \citep{Weisz2014}, where Leo~IV and Hercules
therein are excluded in our analysis because of the large uncertainties in their measured
3D velocities. We also note that we confine ourselves to use the work of \citet{Weisz2014} for the source of
the SF histories to avoid any systematics in their derivations associated with difference methods.

It is evident from Figure~\ref{fig: SFR1} and \ref{fig: SFR2} that the timing of the first crossing
through $r_{\rm vir}$ for each classical dSph, defined as the infall time here, occurs nearly
at the same timing as when the SFR is peaked. For Sculptor, although its first infall did not
cross $r_{\rm vir}$ so the infall time is delayed somewhat, the epoch when it first reached
its pericentric radius occurs nearly at the peak time of the SFR. 

In contrast to these classical dSphs, two UFDs shown in Figure~\ref{fig: SFR3} had the peak of
the SFR well before their infall times. This result is in agreement with that in \citet{Fillingham2019},
which showed that the UFDs shut down their SF prior to the infall to the MW.
However, it is worth remarking that in CVn~I, the second peak of the SFR occurs nearly
at the infall time, whereas in CVn~II, the peak of the SFR is realized at around its first passage
to the MW, although the pericentric radius then did not cross the virial radius of the MW.
This suggests that even in UFDs, their SF histories may be affected by the early
stages of the host, MW halo to some extent.

\begin{figure}[t!]
\centering
\includegraphics[angle=-90, width=80mm]{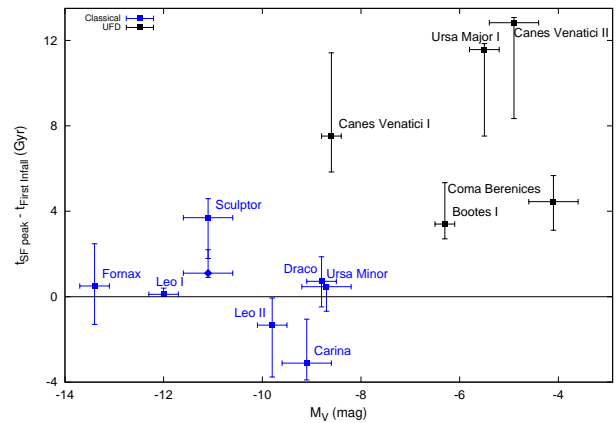}
\caption{
The difference between the time when SFR is peaked and the time of the first infall,
$t_{\rm SF~peak} - t_{\rm first~infall}$ as a function of V-band absolute magnitude, $M_V$,
for the classical dSphs (filled blue squares) and UFDs (filled black squares).
For Sculptor, the filled diamond denotes
the difference between the time when SFR is peaked and the time when it first reached
the pericentric radius.
}
\label{fig: time_diff}
\end{figure}

In Figure~\ref{fig: time_diff}, we show the difference between the 
time when the SFR is peaked and the time of the first infall, $t_{\rm SF~peak} - t_{\rm 
first~infall}$, as a function of V-band absolute magnitude, $M_V$, of each galaxy.
The error bar for the time difference stems from the effect of the range of
the adopted $M_{\rm vir} (z=0)$ on the estimate of $t_{\rm first~infall}$.
For UFDs other than CVn~I and CVn~II, we use the results of {\it HST} observation
by \citet{Brown2014}, which estimate the ages of the dominant old stars in their sample of UFDs.
It is clear that in the classical dSphs, this time difference is confined only
within a few Gyrs. Sculptor shows somewhat a large time difference because its first
infall does not cross within the virial radius of the MW. Instead of using the infall time for this galaxy,
we also plot, with the filled diamond, the difference between the time when the SFR is peaked and
the time when it first reached the pericentric radius, which is now found to be small.
In fact, the time of the first arrival at the pericentric radius occurs just after the
infall time as long as this first infall crosses within the virial radius of the MW.
In contrast, for UFDs, the peak of the SFR occurred much earlier than their first infall to
the MW's virial radius, in agreement with \citet{Fillingham2019}.

This comparison between the orbital evolution and SF history of each classical dSph
shown in Figure~\ref{fig: SFR1} and \ref{fig: SFR2} suggests that not only the
similarity between the infall time and the peak time of the SFR, but also the notable
properties of the subsequent SF activity after the first pericentric passage are
inferred, such that when the pericentric radius of the first infall is small,
the SFR after its peak appears to be considerably reduced.

\begin{figure}[t!]
\centering
\includegraphics[angle=-90, width=80mm]{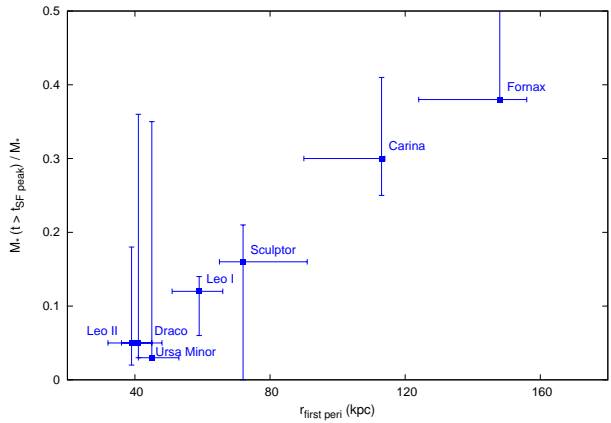}
\caption{
The relation between the fraction of the stars formed after the peak of the SFR
relative to each dSph's current stellar mass, $M_{\ast}$, and the galaxy's pericentric radius
at its first infall to the MW. There is a remarkable correlation such that the smaller
pericentric radius at the first infall leads to the reduction of star formation
after its peak of the SFR.
}
\label{fig: pericenter}
\end{figure}

This property is presented in Figure~\ref{fig: pericenter}, which shows the relation between
the fraction of stars formed after the peak of the SFR relative to each dSph's current
stellar mass, $M_{\ast}$, and the first pericentric radius.
Again, the error bar for the latter quantity stems from the effect of the range of
the adopted $M_{\rm vir} (z=0)$ on the estimate of $t_{\rm first~infall}$.
The clear tight correlation between these quantities is found, such that
the relative amount of stars formed after its peak is reduced
for the smaller pericentric radius at the first infall.
This result is basically consistent with that in \citet{Fillingham2019}, which showed
that the quenching timescale of SF, i.e., the difference between the quenching time,
defined as the look-back time when 90~\% of the current stellar mass is formed, and
infall time is short for the dSphs having high orbital eccentricities.

\section{Discussion and concluding remarks}  \label{sec:discussion}

We have investigated the long-term orbital motions of Galactic dSphs in the course of the 
growing mass of the MW's dark halo over the past 13.5 Gyrs. Their current motions are 
taken from the recent compilation \citep{Riley2019}, which adopts the high-precision measurements
of proper motions from {\it Gaia} DR2. 
We have compared these orbital motions with the SF histories of the dSphs. It is 
found that the infall time of each classical dSph first crossing within the time-varying
MW's virial radius coincides remarkably well with the time when its SFR is peaked.
Also, in the classical dSph whose pericentric radius after the first infall is small, the SFR
afterwards is reduced. Finally, we have confirmed that the formation of stars in UFDs
is finished well before they first enter into the virial radius of the MW's dark halo
as already suggested in previous works. For these UFDs however, we have found a signature
that their earlier SF histories are subject to environmental effects provided by
the early stages of the MW halo to some extent.

The adopted prescription for the form of the Galactic potential and its time dependence given
in Equation~(\ref{eq: NFW}) and (\ref{eq: Mvir}) is admittedly simplistic. For instance,
the accretion process of dark matter onto a host MW-like halo is no longer spherically symmetric
and continuous, but rather anisotropic and sporadic through merging/accretion of many subhalos.
Also, the time evolution of the MW halo and the resultant orbits of the satellites can be
affected by the environment associated with the formation of the Local Group, including
the falling and binding Magellanic Clouds to the Galactic potential
\citep{Bekki2005,Yozin2015,Kallivayalil2018,Pardy2019,Erkal2019,Patel2020}.

To highlight the effect of adopting the Galactic potential on the satellites' orbits, we have
considered the uncertain range of $M_{\rm vir} (z=0)$, which actually dominates the change of
the entire orbital evolution. In addition to these uncertainties, there exist intrinsic measurement errors
in {\it Gaia} DR2's proper motions for each of satellites, which also yield a range of uncertainties
in $t_{\rm first~infall}$. We thus consider the observational errors in the tangential motions of the
satellites presented in Figures \ref{fig: SFR1}, \ref{fig: SFR2}, and \ref{fig: SFR3}, which reflect
the measured proper motions, and find the associated dispersion of $t_{\rm first~infall}$ as
0.48~Gyr for Draco, 1.56~Gyr for Fornax, 1.91~Gyr for Leo~I, 3.41~Gyr for Leo~II, 0.72~Gyr for Sculptor,
3.14~Gyr for Ursa Minor, 1.75~Gyr for Carina, 2.96~Gyr for CVn~I, and 3.25~Gyr for CVn~II. Therefore, 
these uncertainties are not negligible as well in the studies of the satellites' orbits, but it is
found that the timing of their first infall to the Galaxy remains basically unchanged.

Even considering these uncertainties, it is interesting to note that the infall times of Galactic
satellites derived here are generally in agreement of those based on dissipationless cosmological simulations
by \citet{Fillingham2019} (Figure~\ref{fig: infall_time}), suggesting that the adopted simplified
form of the Galactic potential for the calculation of satellites' orbits is not far from
reality. Said that, for a more comprehensive comparison with the SF histories, it will be important to
consider the above other dissipationless processes that we ignore here as well as the effect
of the later baryonic infall that deepens the Galactic potential, and also to refine
the determination of the current MW mass and its distribution \citep[e.g.,][]{Eilers2019,Hammer2020}.

The correlation between the first infall time of a MW satellite and the time of
its maximum SFR has been suggested from recent hydrodynamical simulations of galaxy 
formation, APOSTLE \citep{Genina2019} and Auriga \citep{Simpson2018}. 
Indeed, these simulations show that the very efficient star formation is achieved
when a subhalo containing cold interstellar gas is first approaching to its pericentric radius.
The snapshots of gas densities shown in the \citet{Genina2019}
suggest that gas inside a subhalo is compressed and removed due to ram pressure 
from hot gas in a MW-sized halo and that this effect is strongest around the pericentric 
radius, whereby star formation may be induced from gas compression.
\citet{Simpson2018} from their high-resolution, zoom-in cosmological simulations also show
that the orbital motion of a subhalo having cold interstellar gas within a MW-sized dark halo
affects both the star formation history and the time evolution of its gas fraction in
each subhalo: the epoch when 90\% of the mass of the currently observed stars in a
subhalo is formed, $\tau_{90}$ in their notation, as well as the timing of the reduction
of its gas fraction appears to be well correlated with the first infall time into a host halo.
It is also suggested from their simulations that a smaller pericentric radius at the first infall 
of a subhalo seems to yield a large reduction of its interstellar gas, thereby suggesting
the suppression of subsequent star formation 
\citep[see also the relevant simulation works by][]{Maccio2017,Frings2017,Buck2019}.
These properties are naturally understood
if interstellar gas in a subhalo is efficiently compressed and then removed through
ram-pressure stripping and/or tidal disturbance. Indeed, the presence of a hot gas in
the MW and its effect on satellite galaxies has been suggested and studied from both
the observations of the spectra of background QSOs and theoretical models
\citep[e.g.,][]{Miller2013,Miller2015,Emerick2016}. 
\citet{Fillingham2016} investigated the efficiency of this environmental quenching in detail and
found a rapid stripping mode of gas at low stellar masses of satellites below $10^8$~$M_{\odot}$
as suggested from observations \citep[e.g.,][]{Lin1983,Slater2014,Weisz2015,Wetzel2015,Fillingham2015}.

Finally, star formation activities in UFDs may be entirely determined by the first collapse of 
a subhalo in the early Universe and the subsequent quenching or suppression of star 
formation by reionization of the Universe \citep{Bovill2009,Brown2014}.
Thus, the formation of stars in UFDs is already finished
when they enter into the MW's dark halo. These properties for UFDs are actually
suggested from the ELVIS simulations \citep{Rodriguez2019,Fillingham2019}.

These dynamical effects of the orbits of Galactic satellites, especially the timing of the first infall
in comparison with their star formation histories, can be imprinted in the chemical abundances
of stars in their member stars \citep[e.g.,][]{Koch2006,Kirby2011,Kirby2013} and also the density
distribution of dark matter halos associated with these dSphs through external tides
\citep[e.g.][]{Walker2009,Pena2010,Walker2011,Bullock2017,Hayashi2017,Kaplinghat2019}.
For instance, in Carina, the multiple maxima of the SFR after the first infall seem to occur
at the subsequent pericentric radii, and these phenomena may trigger gaseous infall and
outflow, whereby governing the metallicity distribution of the member stars \citep{Koch2006}.
Also, these star formation activities and associated feedback effects from supernovae
can modify the density profile of a dark halo in dSphs \citep{Read2005,Pontzen2012}.

These chemical and dynamical information of member stars in Galactic dSphs have been biased
toward an inner part of each dSph compared to its nominal tidal radius, so that the global
chemo-dynamical state, especially in an outer part or up to an outer boundary of each dSph,
which is more sensitive to environmental effects, is yet largely unknown.
Extensive spectroscopic data of stars in Galactic satellites out to their outskirts
will be available from Prime Focus Spectrograph (PFS) to be attached to Subaru Telescope
\citep{Takada2014,Tamura2016}. Using this fiber-fed, wide field of view spectrograph,
Galactic Archaeology survey will allow us to get new insights into these subjects and
thus understand the formation histories of the dSphs in the MW.

\acknowledgments

We are grateful to Kohei Hayashi, Evan Kirby, Tadafumi Matsuno and Miho Ishigaki for useful discussion and comments.
We also thank Denis Erkal, Francois Hammer and Tobias Buck for their invaluable comments that help improve the manuscript.  
This work is supported in part by the JSPS and MEXT Grant-in-Aid for Scientific Research (No. 17H01101, 
18H04434 and 18H05437).


\appendix
\restartappendixnumbering

\section{Orbital evolution in a static Galactic potential}

Here, we show all the orbits when the Galactic potential is fixed in the form of Equation~(\ref{eq: NFW})
(blue lines) in comparison with our results shown in Section 3.1 (orange lines).

\begin{figure*}[h!]
\centering
\includegraphics[width=120mm]{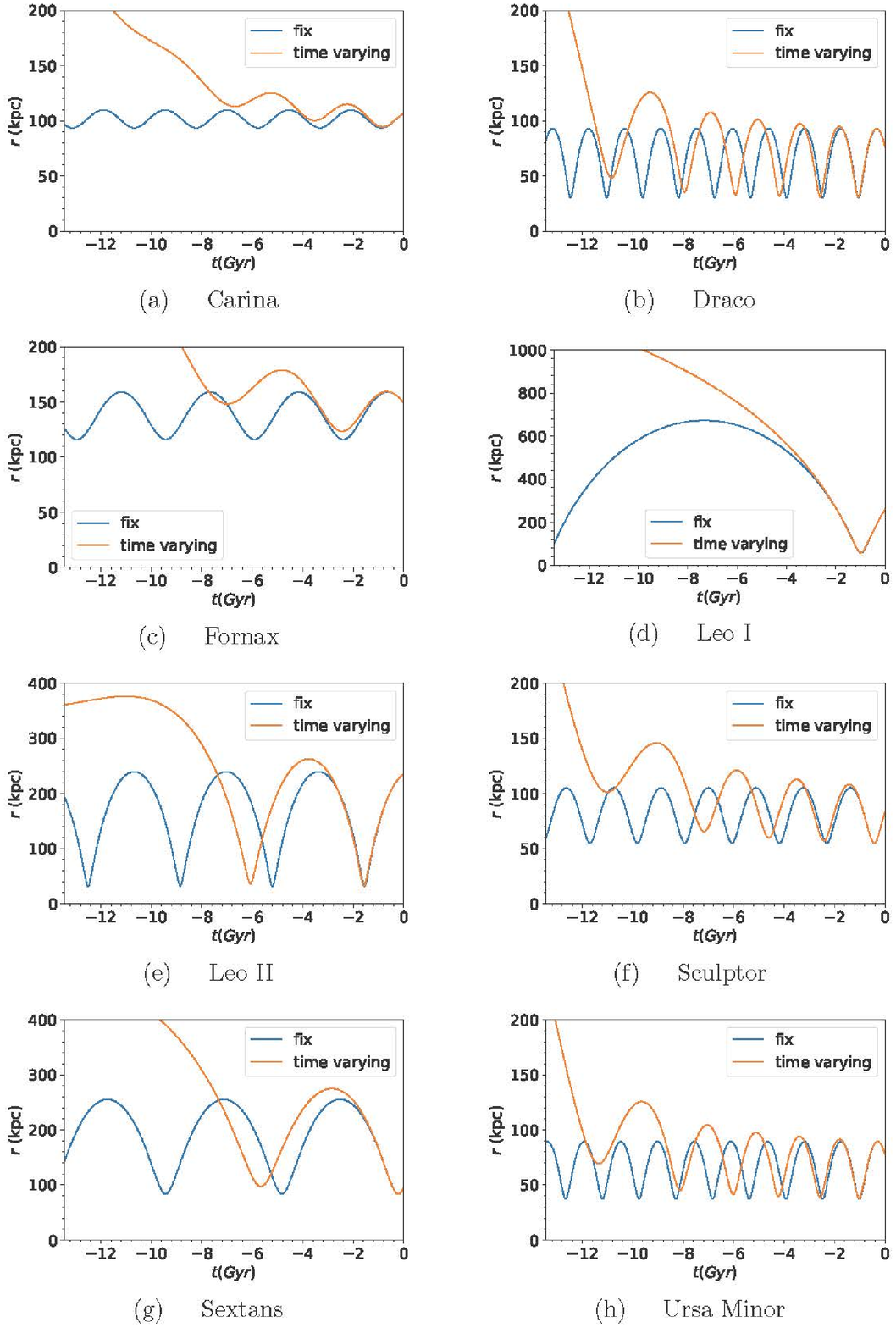}
\caption{
The time evolution of the Galactocentric distance, $r$, of the classical dSphs
when the Galactic potential is static (blue line) in comparison with the time-varying
case (orange line).
}
\label{fig: appendix1}
\end{figure*}
\begin{figure*}[h!]
\centering
\includegraphics[width=120mm]{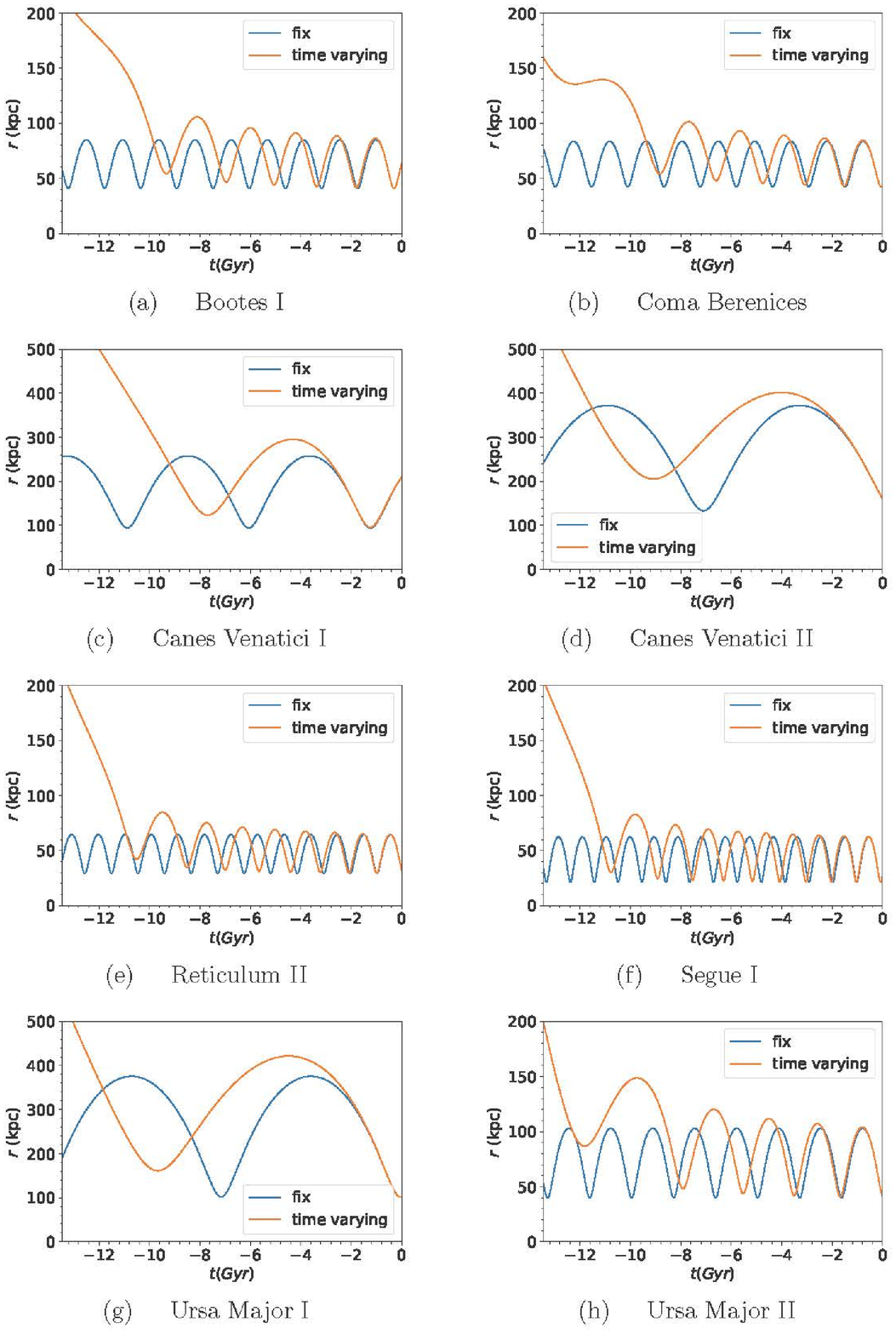}
\caption{
The same as Figure~\ref{fig: appendix1} but for UFDs.
}
\label{fig: appendix2}
\end{figure*}

\end{document}